\documentclass [] {aa} 
\usepackage{epsfig}

\def\kms{\,km\,s$^{-1}$} %kms -1 
\def\ms{\,m\,s$^{-1}$} %ms -1 
\def\vsini{$v\sin i$}
\begin{document}

\thesaurus{06(08.02.4; 08.02.2; 08.09.2 HD209458; 08.16.2)} 
\titlerunning{Spectroscopic transit by the planet orbiting the star 
HD209458} 
\title{Detection of a spectroscopic transit by the planet orbiting the star 
HD209458\thanks{Based on observations collected at the 
Observatoire de Haute-Provence with the echelle 
spectrograph ELODIE at the 1.93m telescope}}

\author{D.~Queloz\inst{1} \and A.~Eggenberger\inst{1} \and M.~Mayor\inst{1} 
\and C.~Perrier\inst{2}, J.L.~Beuzit\inst{2}\and D.~Naef\inst{1}\and 
J.P.~Sivan\inst{3}\and S.~Udry\inst{1}}
\authorrunning{D.  Queloz et al.}

\institute{Observatoire de Gen\`eve, 51 ch.  des
           Maillettes, CH--1290 Sauverny, Switzerland 
      \and
      Observatoire de Grenoble, 414 rue de la Piscine, 
      Domaine Universitaire de St Martin d'Hi\`eres, 38041, France
      \and
      Observatoire de Haute Provence, 
      Saint-Michel l'Observatoire, 04870, France}

\offprints{Didier.Queloz@obs.unige.ch}
%--------------------------------------------------------------------------
\date{Received Jan 2000/ Accepted June 2000} 

\maketitle

\begin{abstract} 
We report the first detection of a planetary transit by spectroscopic 
measurements. We have detected the distortion of the  stellar line profiles 
during a  planetary transit. With the ELODIE spectrograph we took a sequence 
of high precision radial velocities of the star HD209458 at  time of a 
transit of its planet. We detected an anomaly in the residuals of the orbit. 
The shape and the amplitude of the anomaly are modeled as a change of the 
mean stellar line profile resulting from  the planet crossing   the disk of 
the rotating  star. The planetary orbit is in the same direction as the 
stellar rotation. Using  the photometric transit to constrain the timing and 
the impact parameters of the transit, we measure an angle  $\alpha=3.9$ 
[$^0$] between the orbital plane  and the  {\sl apparent equatorial plane}   
as well as a \vsini=$3.75\pm1.25$\kms. With additional constrains on the 
inclination of the star and on the statistics of the line of sight 
distribution,   we  can set an upper limit of 30\,$^0$ to the angle between 
the    orbital plane  and the  stellar  equatorial plane.
\end{abstract}

\keywords{stars: planetary systems - spectroscopic binaries - eclipsing
binaries - individual: HD209458}

\section{Introduction}

A Jupiter mass companion having  a short period orbit was recently detected 
for the star HD209458 by high-precision radial velocity surveys  
(\cite{Henry}, \cite{Mazeh00}). Luckily, the orbital plane of the planet is 
close enough to the line of sight for transits to occur and  be detected by 
photometric measurements (\cite{Charboneau}, \cite{Henry}).
The measurement of the photometric  transit across HD209458 is the first  
independent confirmation of the reality of the  giant planets in short period 
orbits detected by  radial velocity surveys (see \cite{Marcy98} for a 
review). It leads to  the  first  estimate of the radius of a "hot Jupiter" 
planet (or 51-Peg type planet) and it strongly constrains the orbital  
inclination $i_{\hbox{\scriptsize orb}}$.

The crossing of a companion in front of a rotating star  produces a change in 
the line profile of the stellar spectrum. During its transit across the star, 
the companion occults a small area of the stellar disk. If the star is 
rotating, the stellar line profiles  will be distorted according to the 
location  of the planet in front of the stellar disk. 
This phenomenon was already well known from past observations  of eclipsing 
binaries. It was first detected on $\beta$\,Lyrae and Algol systems by 
\cite{Rossiter24} and \cite{McLaughlin24}.  
Recently, it has been suggested by \cite{Schneider00} that a transit by a 
planet could also be detected in the line profile of high signal-to-noise 
ratio stellar spectra for stars with high \vsini. 
But most of the stars with a close-in planet  have low  \vsini~ values and in 
such a case the  line profile distortion  by a transit would be  less than 
1\%  of the width of the lines. This small effect is extremely challenging to 
detect for individual lines, but if a multi-line approach  -- like in  the 
cross-correlation technique -- is used, the mean effect could be  large 
enough to be measured. Actually, any slight distortion of  the stellar line 
profile changes the  radial velocity measured from the Doppler effect, 
similar to the effect of stellar spots on rotating  stars (see 
\cite{Queloz99} for references and details). Current high-precision radial 
velocity measurements probably offer the easiest way to detect a 
spectroscopic transit of a giant planet.

The timing and the amplitude  of the drop of the stellar luminosity observed 
during a planetary transit measure the radius of the planet and its  orbital 
inclination to the line of sight.
The  detection of a spectroscopic transit by radial velocity measurements 
provides a unique  means to estimate the relative inclination ($\alpha$) 
between the stellar equatorial plane projected on the line of sight (called 
hereafter the {\sl apparent equatorial plane})   and the orbital plane,  as 
well as the ascending node of the orbit ($\Omega_p$) on the {\sl apparent 
equatorial plane}.

From a weak friction model (\cite{Hut81}) we find that the tidal effect of a 
short period Jupiter-mass planet on the star is not strong enough to force  
coplanarity.  Comparison between the coplanarization  time and the stellar 
circularization time  indicates that  the alignment time is  100 times longer 
than the circularization time. The stellar circularization time  is of the 
order of a billion years (\cite{Rasio96}).  
Usually one makes the assumption that the orbital plane is coplanar with the 
stellar equatorial plane for close-in planets.
Combined with the  \vsini~ measurement  of the star, this ad-hoc assumption 
is used to  set  an upper limit to the mass of the planet (e.g. 
\cite{mayor95}). The shape of the radial velocity anomaly during the transit 
provides a  tool to test this hypothesis. Moreover, the coplanarity 
measurement is also a way to test the formation scenario of 51-Peg type 
planets.
If the close-in planets  are the outcome  of extensive orbital migration, we 
may expect the orbital plane to be identical to the stellar equatorial plane. 
If other mechanisms such as gravitational scattering played a role, the 
coplanarity is not expected. A review of formation mechanisms of close-in 
planets may be found in   \cite{weidenschiling96} and \cite{lin99}.  

The amplitude of the radial velocity anomaly stemming from  the transit is 
strongly dependent on the star's \vsini~ for a given   planet radius. A 
transit across a star with high \vsini~ produces a larger radial velocity 
signature than across a slow rotator. However it is more difficult to measure 
accurate radial velocities for stars with high \vsini. It requires higher  
signal-to-noise spectra because  the line contrast is weaker.  A star like 
HD209458 with  \vsini~ about 4\kms~  is a good  candidate  for such a 
detection.  With the large wavelength domain of ELODIE (3000\AA) 
approximately 2000 lines are available for the cross-correlation thus only 
moderate signal-to-noise ratio spectra (50-100) are required. 

If we use the  planet's radius derived from the photometric transit,  the 
\vsini~ of the star can be estimated from the measurement of the 
spectroscopic transit.  Unlike spectral analysis, the measurement of the  
\vsini~ provided by the spectroscopic transit is almost independent of the 
accurate knowledge of the amplitude of the spectral broadening mechanism 
intrinsic to the star.  A complete description of  transit measurements is 
given in \cite{Kopal59} for eclipsing binaries and Eggenberger et al. (in 
prep.) for planetary transit cases.

\section{The measurement of the spectroscopic transit}

During the transit, on  November 25th 1999,  we got a continuous sequence of 
15 high precision radial velocity measurements  with the spectrograph ELODIE 
on the 193cm telescope of the Observatoire de Haute Provence 
(\cite{Baranne96}) using the simultaneous thorium setup. The following night  
we repeated the same sequence, but off-transit this time, in order to check 
for any instrumental systematics possibly stemming from the relative low 
position on the horizon. For both nights the sequence was  stopped when a 
value of two airmasses  was reached. The ADC (atmospheric dispersion 
corrector)   does not correct efficiently at higher airmass.

As usual for ELODIE measurements, the data reduction was made on-line at the 
telescope. The radial velocities have been  measured by a cross-correlation 
technique with our standard binary mask and Gaussian fits of the 
cross-correlation functions (or mean profiles) (see \cite{Baranne96} for 
details).

\begin{figure}
\psfig{width=\hsize,file=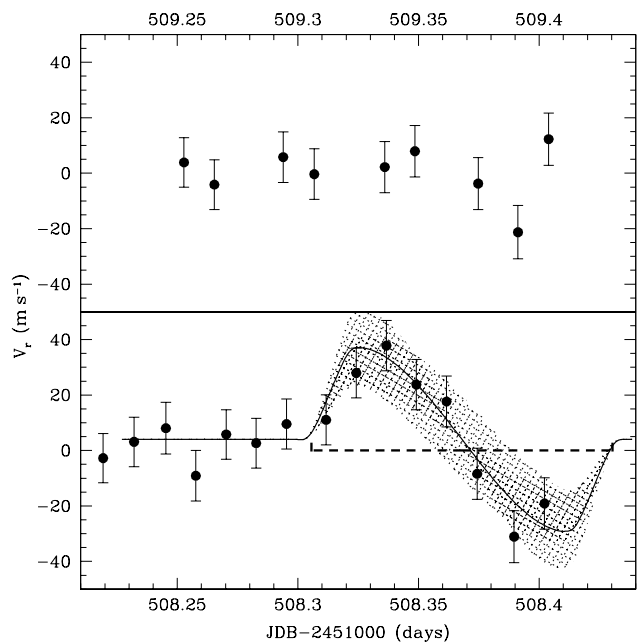}
\caption[]{Two sequences of radial velocity residuals for the star HD209458 
taken at the same time during  the night   but one day apart. The data are 
corrected for the orbital motion of the planet with the \cite{Mazeh00} 
ephemeris.  {\bf Top} out of the planetary transit, the residuals agree with 
random error. {\bf Bottom} during the planet  transit an anomaly is detected. 
The duration of the photometric transit is indicated by the thick dashed 
line. Notice the good timing agreement between the beginning of the  
photometric transit and the beginning of the radial velocity anomaly.
Our best  model  of the radial velocity  anomaly (see below) is superimposed 
on the data (solid line). The $1\sigma$ confidence level of the anomaly model 
is illustrated by the dotted area}
\end{figure}

The residuals from the spectroscopic orbit of HD\,209458  (\cite{Mazeh00})  
are displayed for two selected time spans in Fig.\,1. During the transit an 
anomaly is observed  in the residuals. The probability to be a statistical 
effect of  a random noise distribution is $10^{-4}$ ($\chi2=53.4$). The 
second night with the same timing sequence no significant deviation from 
random residuals is observed. Note that the usual 10\ms long-term 
instrumental error has not been added to the photon-noise error since  the 
instrumental error is negligible on this time scale, accordingly with the 
40\% confidence level measured for the non-signal model during the 
off-transit night.

\section{Modeling the data}

In Fig.\,2  the geometry of our model is illustrated. The  orbital motion is 
set in the same direction as  the stellar rotation. This configuration  
actually stems from the  transit data themselves:   the radial velocity 
anomaly   first has a positive bump and then a negative dip. This tells us 
that the planetary orbit and the stellar rotation share the same direction 
whatever the geometry of the crossing may be (direct orbit). 

\begin{figure}
\psfig{width=\hsize,file=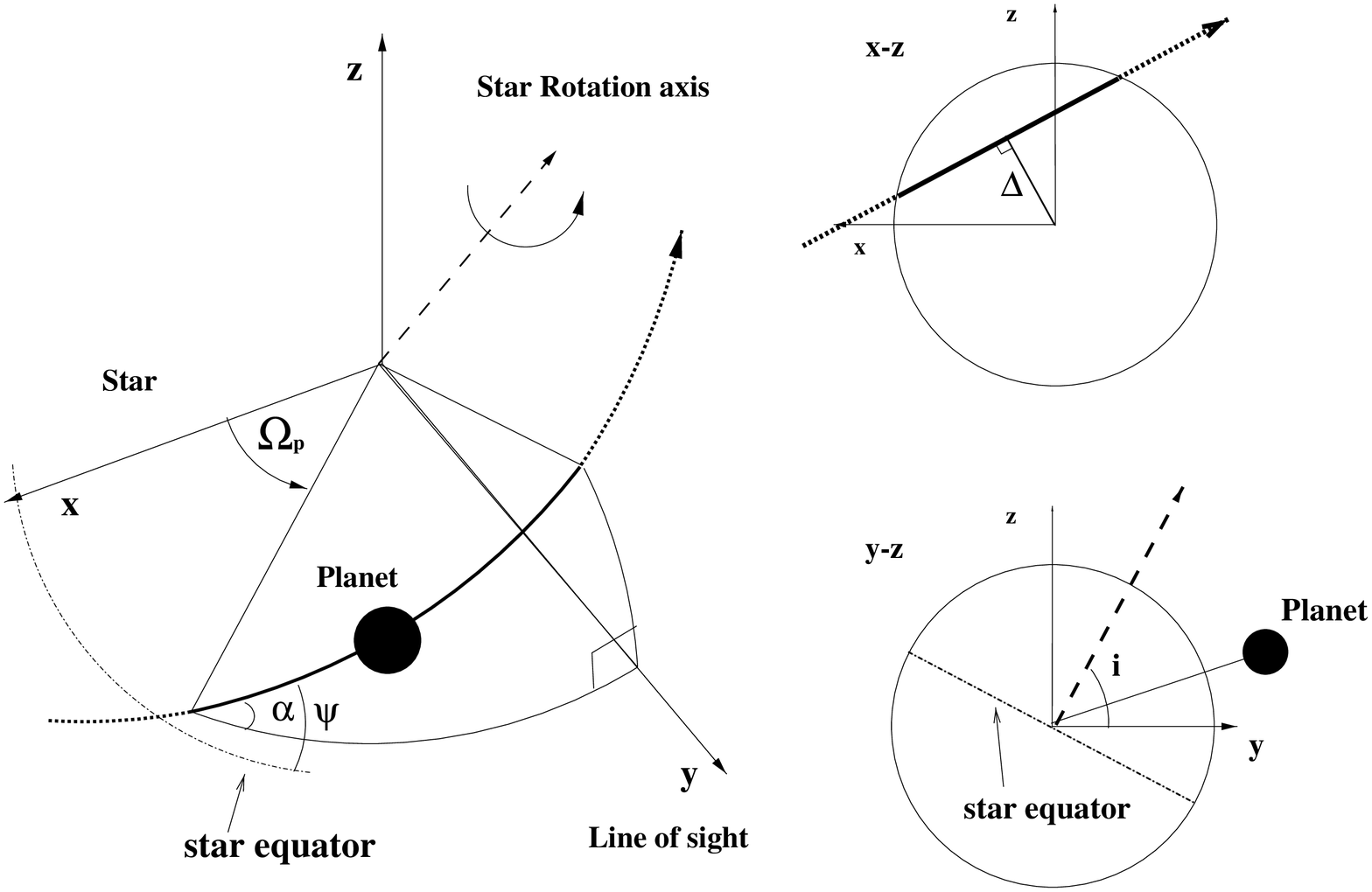}
\caption[]{Illustration of the geometry of our model. The $(x-y)$ plane is 
called {\sl apparent equatorial plane}. The y-axis is the  line of sight.
Notice the definition of the angle $\alpha$  and the ascending node 
$\Omega_p$ on the  {\sl apparent equatorial plane}. The angle between the 
star's equatorial plane and the planetary plane is described by $\psi$. On 
the $(x-z)$ plane the thick line represents the orbital trajectory projected 
on the stellar disk.  The stellar rotation axis (dashed line)  is seen  in 
the plane $(y-z)$ with its angle $i$.}
\end{figure}

In this paper we decided to restrict our analysis to the measurement of three 
parameters: the angle $\alpha$ between the orbital plane and the {\sl 
apparent equatorial plane}, the \vsini~ of the star  and the ascending node 
$\Omega_p$. The ascending node is taken positive in the direction of the 
star's rotation and equal to  $90^0$ when crossing the line of sight axis.
The timing of the transit  and the impact parameter $\Delta$ (shorter 
projected distance between the transit trajectory and the star center) are 
set by the results of the photometric transit. The Hipparcos measurements of 
the transit (\cite{Robichon00}) constraint a very accurate   orbital period 
of the system. Combined with the mid-time of the transit by \cite{Mazeh00}, 
we know the observed transit mid-time with an accuracy of 5\,minutes. 
More radial velocity data at high  precision would be   required to make a 
full adjustment of the transit (timing and geometry).  In Table\,1 we list  
the results   from the spectroscopic orbit and the photometric transit that 
were used to constrain our model adjustments.

\begin{table}
\caption[]{Star, planet and  orbit parameters of our model}
\begin{tabular}{llr}
\hline
\multicolumn{2}{c}{fixed parameters}\\
\hline
\noalign{\smallskip}
$P$                              & & $3.524739\pm1.4\cdot10^{-5}$\,d\\
$a$                              & & $0.047\pm0.001$\,AU\\
$e$                               && $0$\\
$T_c^{\diamond}$                           &  & $2451508.368\pm0.0032$\\
$R_\star$                       & & 1.2($\pm0.1$)\,R$_\odot$\\
$R_{\hbox{\scriptsize\rm planet}}$&& 1.40($\pm0.17$)\,R$_J$\\
$\Delta^{\ddagger}$                          &&$0.569\pm0.004$\,R$_\star$\\
\noalign{\smallskip}
\hline
\multicolumn{2}{c}{Our best solution }\\
\hline
\noalign{\smallskip}
\vsini                          &  & $3.75\pm1.25$ \kms\\
$\alpha$                         &   & $\pm3.9^{+18}_{-21}$ [$^0$]\\
$\Omega_p^{\dagger}$                     &  for $\alpha=3.9$ & $0$ [$^0$]\\
                                         &  for $\alpha=22$ & $100$ [$^0$]\\
                                         &  for $\alpha=-25$ & $81$ [$^0$]\\
\noalign{\smallskip}
\hline
\end{tabular}
\\ Orbit and transit data are from Mazeh et al. (2000) excepted for the 
period 
   which is from  Robichon \& Arenou (2000)
\\  ($^{\diamond}$) computed with $P$ of Robichon \& Arenou (2000) and $T_c$ by Mazeh 
et al. (2000)
\\  ($^\ddagger$) derived from the orbital inclination angle $86.1^0\pm1.6^0$
\\  ($^\dagger$) $\pm 180$\,[$^0$] undefined
\end{table}
 
In our model we consider a spherical star in uniform rotation. The star is 
divided  into 90000 cells. A Gaussian shape cross-correlation model with 
$\sigma_0=3.0$\kms width is used to model the  mean individual  spectral line 
of each cell where the center of the Gaussian is equal to the mean radial 
velocity of the cell. The effect of the $\sigma_0$ value on the \vsini~ 
estimate is negligible: with $\sigma_0=2.5$\kms~ we would only  increase the 
\vsini~ measurement by 0.1\kms.

The integration is made by summation of the cells free of the planet along 
the line of sight. A linear limb darkening weighting ($B_\mu=1-\epsilon 
(1-\mu)$) with $\epsilon=0.6$ is used in the sum. The planetary orbit is 
circular. We divide the transit into 50 phase steps for computing the radial 
velocity anomaly of the spectroscopic orbit residuals. 

To illustrate the effect of the transit geometry on the radial velocity 
anomaly during the transit, we display in Fig.\,3 the  anomaly  expected  for 
three geometric configurations and two different \vsini~ values. We see   
that the  amplitude of the anomaly is driven by   the \vsini~  value of the 
star. The radial-velocity symmetry  of the anomaly of the spectroscopic orbit 
residuals    (from the mid-transit) for $\Delta>0$  impact parameter  depends 
on  the angle $\alpha$ between the stellar {\sl apparent equatorial plane} 
and the orbital plane.   Interestingly enough, the  $\alpha$ parameter is not 
tied to the \vsini~ measurement. The two parameters are uncorrelated. In 
Fig.\,3 we also see that the mirror trajectories along the {\sl apparent 
equatorial plane} make similar radial velocity anomalies because $\Omega_p$ 
is $\pm 180$\,[$^0$] undefined.  Nevertheless, the value of $\Omega_p$ can be 
used to make the distinction between cases a and  b from  Fig.\,3.

\begin{figure}
\psfig{width=\hsize,file=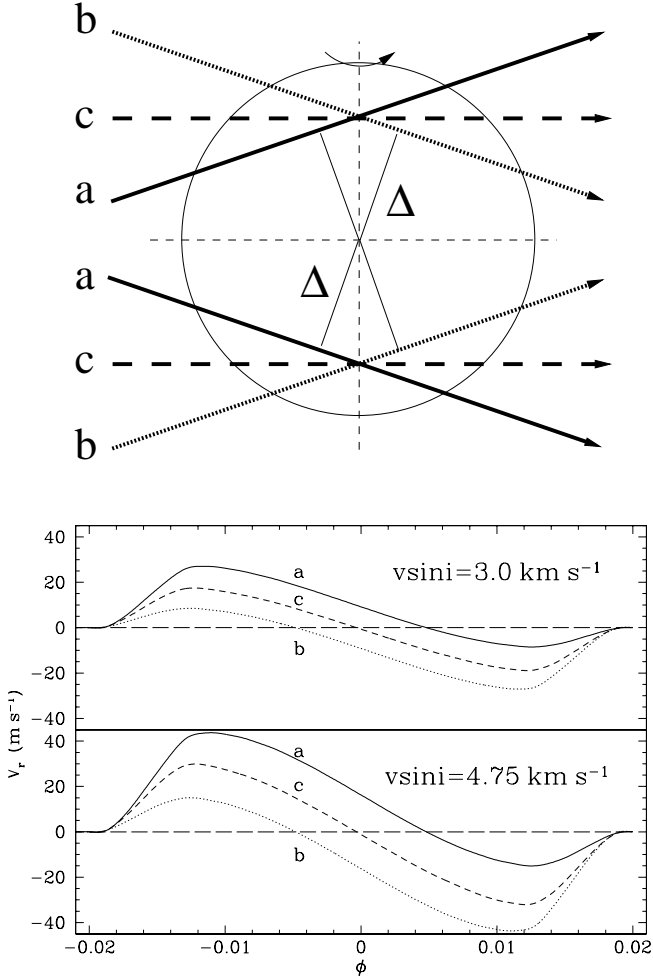}
\caption[]{Illustration of the effect of the geometry  of the transit on the 
radial velocity  anomaly for HD209458: {\bf Upper panel}. Geometry of the 
crossing for 3  cases (+3 symmetrical cases): (a) $\alpha=20$\,[$^0$] and 
$\Omega_p\in[0;90]\cup[180;270]$; (b)   $\alpha=20$\,[$^0$] and 
$\Omega_p\in[90;180]\cup[270;360]$; (c)  $\alpha=3.9$\,[$^0$] (horizontal 
trajectory).
{\bf Lower Panel}. Radial velocity anomalies for the geometries displayed 
above for ({\bf top})  \vsini=3\kms  and ({\bf bottom}) \vsini=4.75\kms}
\end{figure}

The comparison of the  transit model with the data is made by $\chi^2$ 
statistics with  the velocity offset, the star's \vsini, and   $\alpha$ as 
free parameters. To distinguish the geometry of the two  trajectories  that 
have similar $\alpha$ angle but different $\Omega_p$ value, we arbitrarily 
set $\alpha<0$ when 
$\Omega_p$ value is between 0 and 90 (or 180-270) 
and $\alpha>0$ when 
$\Omega_p$ value is between 90 and 180 (or 270-360). These are respectively 
the a and b trajectories  illustrated on Fig.\,3.

The best set  of $\alpha$-\vsini~ solutions is listed in Table.\,1    with 
$1\sigma$ confidence levels. Following our  definition of $\alpha$, any angle 
with $3.9<\alpha<-3.9$\,[$^0$] is impossible given our impact parameter 
$\Delta=0.569$\,R$_\star$.  Our best fit is reached when $\alpha=\pm3.9$ and 
$\Omega_p=0$. It corresponds to  a transit trajectory parallel to the 
{\sl apparent equatorial plane}. 

The  uncertainties on our best solution listed in Table.~1 do not include 
systematics stemming from errors in fixed parameters of the model. A 
($\pm0.17\,R_J$) change of the planet radius would lead to a change of  the 
\vsini~ by $\pm$0.75\kms. Actually if the planet has indeed a larger radius  
we overestimate the \vsini~ value. The uncertainty on the star radius has a 
weaker  effect on the \vsini~ measurement than the uncertainty on the planet 
radius. A  change of the star radius by $\pm0.1\,R_\odot$ would make the 
\vsini~ change   by 0.3\kms only.

\section{Discussion}

We have successfully demonstrated  the detection of a planetary transit  
using a time sequence of stellar spectra when the stellar line profile is 
distorted by the crossing of the planet and changes  the radial velocity 
measurement of the star.
We have detected an anomaly in the residuals of the radial velocity orbit of 
HD\,209458 at the time of the transit. This anomaly has been modeled with 
high confidence  as the effect of the planetary transit. The  data suggest an 
orbital trajectory parallel to the {\sl apparent equatorial plane}.

The measurement of an $\alpha$  different from zero and an ascending  node 
off the plane of the sky  ($\Omega_p\neq0$)  would  indicate  deviation from 
the coplanarity between the orbital plane and the stellar equatorial plane. 
In our case we can only argue that  no evidence for non-coplanarity is found. 
However additional arguments  can set some limits on the coplanarity level. 
From the statistics only, it would be "bad luck" to observe a  non-coplanar 
system in a configuration such as $\alpha\approx0$ and $\Omega_p=0$. This 
would only  happen  for a very small set of system orientations. 
We computed the  probability of finding  $\alpha$ smaller than $\alpha_{max}$ 
for a system with an angle $\psi$  between  its orbital plane and  the 
equatorial plane (see on Fig.\,2 for an illustration of the $\psi$ angle). 
First, we have determined the relation between $\alpha$ and $i$ for various 
$\psi$ configurations. Simply  said we have computed for each configuration 
all the different spectroscopic signals  that  extra-terrestrial observers 
looking at the system from  different points of view could see the same 
photometric  transit. Then  we have  generated  series of random numbers $i$ 
by recalling that the probability of seeing a system with a small $i$ is 
smaller
than the probability of seeing a system with a large $i$. For each of these
values we have used the relation between $\alpha$ and $i$ to calculate the
corresponding statistical distribution of $\alpha$. Finally  the cumulative 
distribution of $\alpha$ is shown on Fig.\,4.  With $1\sigma$ confidence 
level  we   can rule out a $\psi$ angle larger than  30 degrees.

To get a complete description of the coplanarity between the stellar 
equatorial plane and the orbital plane, measurement of the  angle $i$ is 
required. Ideally  $i$ can be estimated from the rotation rate of the star.
HD209458 has a quiet chromosphere ($R_{HK}=-4.9$ by \cite{Henry}) but  
significant  spectral line rotation broadening (\vsini) is detected. The 
spectroscopic \vsini~ measurements have a  weighted mean value of $3.7\pm0.5$ 
\kms (measurements by \cite{Mazeh00} and Marcy et al. private comm.) in 
agreement with the value measured in this work.
The low chromospheric HK value suggests a rotation period of at least 
17\,days (\cite{Noyes84}). With the  \vsini~ and the  period estimate  we 
find $i\tilde{>}60^0$.

\begin{figure}
\psfig{width=\hsize,file=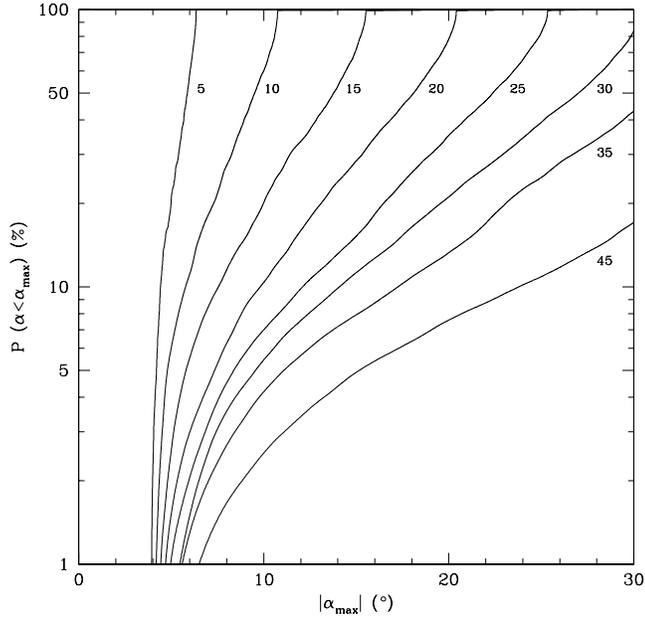}
\caption[]{Probability to measure  an angle $\alpha$ for various 
configurations. The  labels of curves describes the angle $\psi$ ($^0$) 
between the   equatorial plane and the orbital plane} 
\end{figure}

The statistical result  on the geometry distribution of the system added to 
the constraint on $i$ provides two arguments refuting  strong non-coplanarity 
for this system. 
Better measurements of the spectroscopic transit are needed to get stronger 
constrains. Simulations show that for a coplanar system with the same radial 
velocity accuracy but 4 times more data spread over the whole transit 
duration, the error bars on the measurement of  $\alpha$ would be small 
enough to conclude that $\psi<10[^0]$ with $1\sigma$ confidence level.

\begin{acknowledgements}
We are gratefull to the 193cm-telescope staff of Observatoire de 
Haute-Provence and in particular the night assistants for their efforts and 
their efficiency. We thank our referees   G. Henry and F. Fekel for  useful  
comments and suggestions about this work and Yves Chmielewski for his help  
for the spectral synthesis. We acknowledge the support of the Swiss NSF 
(FNRS). 
\end{acknowledgements}

%---------------------------bibliography---------------------------

\newpage

\end{document}